\documentclass[letterpaper]{sfchem}
\usepackage{graphicx}

\begin{document}

\title{Chemical changes during star formation: high- vs.\ low-mass YSOs}

\author{Ewine F.\ van Dishoeck\inst{1}}
  \institute{Leiden Observatory, P.O.\ Box 9513, 2300 RA Leiden,
             The Netherlands}
\authorrunning{van Dishoeck}
\titlerunning{High- versus low-mass YSOs}

\maketitle 

\begin{abstract}
Recent observational studies of high- and low-mass YSOs at
 (sub)millimeter and infrared wavelengths are reviewed, and chemical
 diagnostics of the different physical components are summarized.
 Methods for determining the temperature, density and abundance
 profiles in the envelopes are outlined, and are illustrated for one
 high-mass and one low-mass YSO. The combination of (sub)millimeter
 and infrared data gives a nearly complete chemical inventory of the
 gas and solid state material.  In both high- and low-mass YSOs, the
 chemical characteristics are dominated by freeze-out in the cold
 outer part of the envelope and evaporation of ices in the warm inner
 part. Abundance jumps of factors of $\sim$100 in selected molecules
 are found in the warm gas for both types of objects. Potential
 differences include (i) the complex hot core chemistry, which has
 been observed so far only for high-mass YSOs; (ii) the high level of
 deuterium fractionation seen only in low-mass YSOs; (iii) the effects
 of internal or external UV and X-rays; (iv) the relative importance
 of shocks versus thermal heating of the envelope; and (v) the
 importance of geometrical effects.  Prospects for future
 observational facilities are mentioned.

\keywords{ISM: molecules -- ISM: chemistry -- Star formation: high-mass -- Star
formation: low-mass}

\end{abstract}

\section{Introduction}
  
The formation of low- and high-mass stars is accompanied by large
changes in the temperature and density structure in their surrounding
envelopes.  The chemistry responds to these changes by opening up
different chemical routes (e.g., van Dishoeck \& Blake 1998, Hartquist
et al.\ 1998, Langer et al.\ 2000). For example, in the coldest
($\leq$10~K), densest regions the time scales for freeze-out are so
short that most molecules stick onto the grains.  Close to the young
protostars, the grains are heated to at least 100~K resulting in the
evaporation of ices which can subsequently drive a rich gas-phase
chemistry. At even higher temperatures ($\geq$230~K), most of the
oxygen is driven into water and a different chemistry ensues. In this
paper, a brief overview is given of recent systematic studies of the
chemistry in the envelopes around young stellar objects (YSOs), with
special attention to the similarities and differences between low- and
high-mass sources.

Progress in this field has been driven by several new developments in
the last decade. First, submillimeter observations routinely sample
smaller beams ($\sim 15-30''$) and higher critical densities ($\geq
10^5$ cm$^{-3}$) than earlier millimeter studies, assuring less
confusion with the lower-density cloud material. Bolometer arrays now
readily provide maps of the continuum emission from cold dust,
constraining the mass and the density structure of the envelope.  High
spatial resolutions down to $\sim 1''$ can be obtained with
interferometers, although chemical data on YSOs are still limited. For
comparison, most low-mass objects observed with these facilities are
located at $\sim$150 pc, where 20$''$ corresponds to $\sim$3000 AU
---the typical size of an envelope--- and $1''$ to 150 AU ---the
typical size of a circumstellar disk. The observed high-mass YSOs
have distances of 2--4 kpc, so that much larger scales are
probed. Nevertheless, it is possible to derive information on smaller
scales by careful modeling of the line emission (see \S 2).

Second, infrared observations have matured and now provide essential
complementary information to the millimeter data. Not only gas-phase
molecules, but also ices, PAHs and atomic fine-structure lines can be
observed at mid-infrared wavelengths -- unique probes of different
evolutionary stages (e.g., van den Ancker et al.\ 2000a,b, Nisini et
al.\ 2002).  The single-dish submillimeter emission is more sensitive
to the colder outer envelope, whereas the infrared absorption is
weighted toward the warmer inner region. Together, they allow a nearly
complete inventory of the gas-phase and solid-state material and can
firmly establish abundance gradients across the envelope.

Third, the tools for the analysis of the observational data have
improved substantially. Continuum and line radiative transfer codes
are being extended from 1-dimension (1D) to 3-dimensions (3D) (e.g.,
Juvela 1997, Wolf et al.\ 1999). 
More importantly, the 1-D codes have been tested
critically against each other, establishing their accuracy (van
Zadelhoff et al.\ 2002).

Several systematic studies of the chemistry in samples of low- and
high-mass YSOs have been carried out in the last five years or are in
progress (e.g., Hatchell et al.\ 1998a,b, Hogerheijde et al.\ 1999,
van der Tak et al.\ 2000a,b, Lahuis \& van Dishoeck 2000, Shah \&
Wootten 2001, Buckle \& Fuller 2002, J{\o}rgensen et al.\ 2002,
Boonman et al.\ 2003b,c, Johnstone et al.\ 2003, Maret et al., in
prep.).  Questions to be addressed with these surveys are: (i) What is
the physical structure of the envelope ($T(r)$, $n(r)$)?  Do continuum
and line data give the same results? (ii) Which molecule/line traces
which component? Can one uniquely distinguish envelope, outflow, disk
and surrounding low-density cloud? (iii) How does the chemical
composition change between different sources? Is this related to
evolution or do other parameters play a role? (iv) What are the
differences between high- and low-mass YSOs? (v) What do the data tell
us about basic chemical processes, in particular gas-grain
interactions, grain-surface chemistry and high temperature gas-phase
chemistry?

In this paper, a brief review and progress report of recent results is
given, focussing on items (i), (iii) and (iv).  In \S 2, the analysis
methods are outlined in more detail. In \S 3, the results on high-mass
YSOs are summarized and illustrated by discussing one source in detail
(AFGL 2591).  In \S 4, the same recipe is followed for low-mass YSOs,
using IRAS 16293-2422 as the example. In \S 5, hot core and envelope
models are summarized. Similarities and differences between high- and
low-mass YSOs are discussed in \S 6, followed by a summary in \S 7.
For more extensive reviews of older work, see van Dishoeck \& Blake
(1998), van Dishoeck \& Hogerheijde (1999) and van Dishoeck \& van der
Tak (2000).

\section{Analysis methods}

Tradionally, chemical abundances have been derived using simple tools,
such as rotation diagrams assuming a constant excitation temperature
$T_{\rm ex}$ or statistical equilibrium calculations using an escape
probability for a single temperature $T$ and density $n$. In both
cases, comparison with a `standard' molecule such as CO is needed to
derive the abundance with respect to H$_2$. It is well known that this
method can lead to abundances that are in error by more than an order
of magnitude.  For example, the use of $n=10^5$ versus $10^6$
cm$^{-3}$ for the interpretation of the HCN $J$=4--3 line with $n_{\rm
cr}\approx 10^7$ cm$^{-3}$ gives a factor of 10 difference in derived
HCN abundance. Densities are known to range from $10^4$ to $>10^7$
cm$^{-3}$ across protostellar envelopes, so which density is the
appropriate one for this line?  Neglect of beam dilution and optical
depth effects can result in similar or even larger errors. For
example, the emission from the complex organic molecules may originate
in the inner $1''$, much smaller than the observational beam of $\sim
15''$.

\begin{figure}[ht]
\includegraphics[width=7cm,angle=-90]{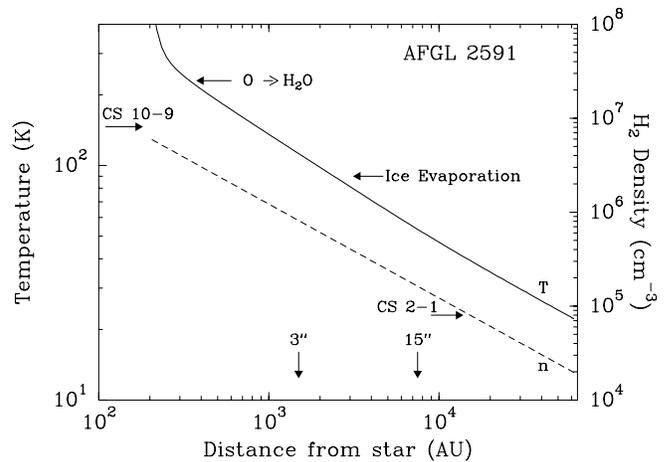}
\caption{Derived temperature and density structure for the envelope
around the massive YSO
AFGL 2591 (L=$2\times 10^4$ L$_{\odot}$, $d$=1 kpc). 
The critical densities of the CS 2--1 and 10--9 lines are
indicated, as are typical single-dish beam sizes ($\sim 15''$) 
and interferometer resolutions ($\sim 3''$). Note that the different
chemical regimes are not resolved in single-dish data
(based on: van der Tak et 
al.\ 1999).}
\label{gl2591}
\end{figure}

\begin{figure*}[t]
\vspace{-0.8cm}
\includegraphics[width=11.0cm,angle=-90]{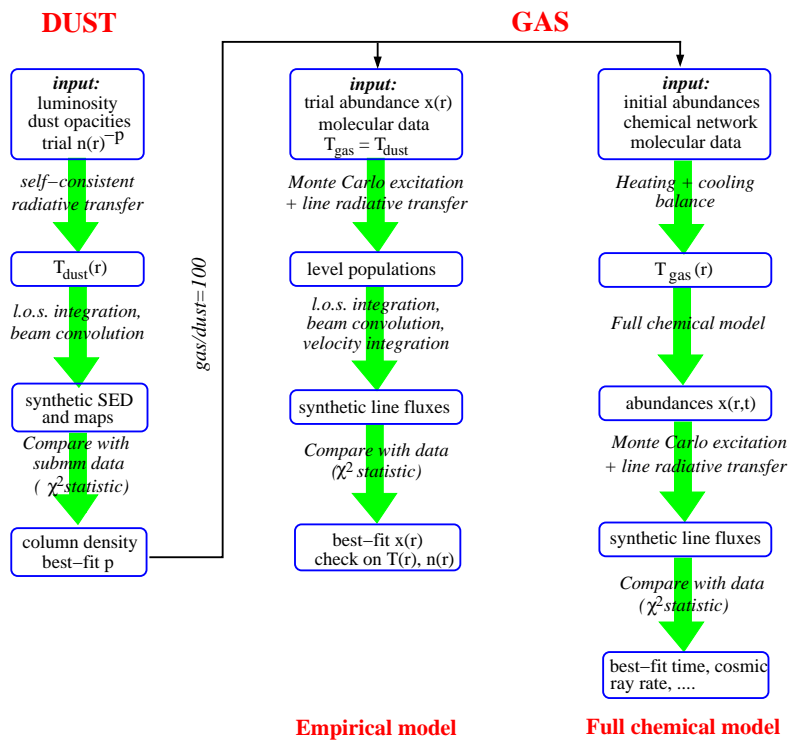}
\caption{Overview of methods used for constraining the physical and
chemical structure of YSO envelopes.  The chemical analysis can proceed in
two ways. In the `empirical' analysis, $T_{\rm gas} = T_{\rm dust}$ is
assumed and the abundance of each species is derived by a $\chi^2$-fit
to the observed fluxes.  In the `full chemical model' analysis, the
physical structure is coupled with a detailed chemical network and
$T_{\rm gas}$ and the abundances of each species are calculated
explicitly at each radius as a function of time. The emerging fluxes
are then calculated for each molecule and compared with observations
for consistency. The basic model parameters (e.g., cosmic ray
ionization rate, time, initial abundances) are adjusted to obtain the
best fit (based on: van Dishoeck \& van der Tak 2000, Sch\"oier et
al.\ 2002, Doty et al.\ 2002).}
\vspace{-0.3cm}
\label{models}
\end{figure*}

The modern analysis method starts with a physical model of the source,
constrained by observations. The excitation and radiative transfer are
subsequently calculated through this model and the resulting sky
brightness distribution is convolved with the beam profile, or, in the
case of interferometer data, analyzed with the same spatial
filtering. The abundance of the molecule is then adjusted until
agreement with the observational data is reached (see Fig.\ 8 of van
Dishoeck \& Hogerheijde 1999 for outline). To constrain the physical
models, two approaches are taken. Both assume a 1D spherically
symmetric model with a power-law density profile ---the simplest
realistic source model--- and calculate the dust temperature profile
through the envelope explicitly.  In the first method, lines of the
same molecule (e.g., CS 2-1, 3-2, 5-4, 7-6, 10-9) are used to
constrain the power-law index of the density structure (e.g., Zhou et
al.\ 1994, van der Tak et al.\ 2000a) (see Figure~\ref{gl2591}).  The
critical asumption is that the abundance of the molecule does not vary
with position in the envelope. The second, now more robust approach
uses the observed dust continuum maps and spectral energy distribution
(SED) to constrain the density profile (see Figure~\ref{models})
(e.g., Shirley et al.\ 2002, J{\o}rgensen et al.\ 2002).  For each
trial density structure, the dust temperature is calculated and the
preferred power-law index is determined by a $\chi^2$-fit to the
observational data. This method assumes that the dust emissivity does
not vary across the envelope.

The accuracy in derived power-law index is typically $\pm$0.2 and the
density gradient is generally smooth, with little evidence for
clumping.  Within the inner envelope out to a few thousand AU, various
analyses of the same sources usually agree well. The major differences
are in the outer part, especially in the extent of the envelope and
its merging with the surrounding cloud. Also, external UV radiation
can increase the outer dust temperature. Deviations from spherical
geometry due to e.g.\ outflow cavities may affect the optical depths
at infrared wavelengths by up to a factor of three.

Once the physical structure is fixed, the line emission of different
molecules can be calculated to determine the abundances, assuming a
gas/dust ratio of 100 and a gas temperature that is equal to the dust
temperature. The latter assumption is known to fail in the outer,
lower density part of the envelope, where the gas temperature can
either fall below the dust temperature in the absence of external
heating (e.g., Doty \& Neufeld 1997) or be above the dust temperature
in the case of external ultraviolet radiation (e.g., Kaufman et al.\
1998). Test models which calculate the gas temperature explicitly
generally give very similar results for the inferred abundances, so
that this difference can be neglected to first order. For small beams
and high-frequency lines, the differences in extent of the model
envelopes do not affect the abundance analyses.

In the baseline models, the abundance of the molecule is assumed to be
constant with envelope radius and is the single parameter to be fitted
to the data. If lines originating from a large range of energies are
available, abundance {\it profiles} can be constrained.  The simplest
assumption here is a `jump' in the abundance at a temperature where a
chemical change is expected, for example around 100~K where
evaporation of ices occurs.  H$_2$CO, CH$_3$OH and SO$_2$ are three
well-known examples of molecules for which lines with upper level
energies ranging from 10 to 300~K are commonly observed in the
150--370 GHz atmospheric windows and for which such abundance jumps
have been inferred (e.g., Ceccarelli et al.\ 2000, van der Tak et al.\
2000b, Sch\"oier et al.\ 2002).  The free parameters are now the inner
and outer abundances and the location of the jump. The latter
parameter is usually taken to be fixed at the radius where $T=90$~K.
Accurate collisional rate coefficients for the various molecules are
an essential ingredient for these analyses (e.g., Sch\"oier et al.,
this volume). Such data are often lacking for more complex molecules,
where LTE excitation must be assumed.

The scheme discussed above describes the `empirical' modeling to
derive chemical abundances directly from observations. Alternatively,
one can couple chemical models with the physical structure derived
from observations to calculate the abundances explicitly with radius
and time in a `full chemical model'. The radiative transfer for the
different molecules then provides line intensities which can be
compared with observations for consistency. The parameters to adjust
in these models are those that enter the chemical models, such as the
cosmic-ray ionization rate, the initial elemental and molecular
abundances, and the time since the abundances were last reset by some
physical event (e.g., Doty et al.\ 2002).

\section{High-mass YSOs}

\subsection{Recent observations}

A number of recent submillimeter line surveys have highlighted the
large variations in chemical composition toward different YSOs. Toward
Orion-KL, the recent 650 and 850 GHz surveys illustrate the
chemical richness of this source, even at high frequencies (Schilke et
al.\ 2001). Observations toward different positions have
shown large chemical variations, for example in oxygen and
nitrogen-bearing molecules between the `hot core' and `compact ridge'
(e.g., Sutton et al.\ 1995) and in chemical complexity with the
line-poor Orion-S position.  Such differences are also reflected in
infrared spectra toward the Orion-KL region, e.g., by selective
absorption or emission of species like HCN, C$_2$H$_2$ and CO$_2$
(Evans et al.\ 1991, Boonman et al.\ 2003a).  For Sgr B2, the spectra
toward the N, M and NW positions are dominated by organic molecules,
sulfur-bearing species and simple species including carbon chains,
respectively (Nummelin et al.\ 1998, 2000).  Similar characteristics
are seen towards the W3(H$_2$O), W3 IRS5 and W3 IRS4 positions
(Helmich \& van Dishoeck 1997). These variations have
been interpreted as being due to differences in time scales since
evaporation of the ices and variations in initial ice compositions
(e.g., Charnley et al.\ 1992, Rodgers \& Charnley 2001, see \S 5).

\begin{figure}[ht]
\includegraphics[width=5cm,angle=-90]{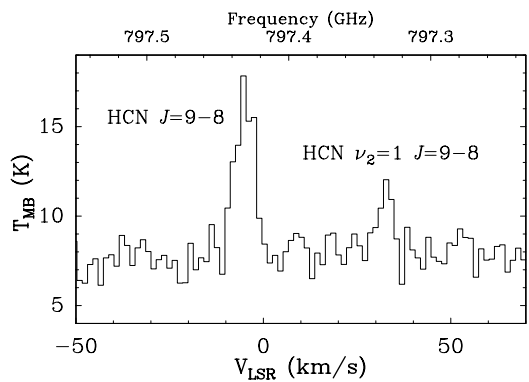}
\caption{Detection of the HCN $\nu_2$=0 and $\nu_2$=1 $J$=9--8 line 
toward AFGL~2591 with the MPIfR/SRON 800 GHz receiver on the JCMT. The
vibrationally-excited emission indicates a jump of a factor of $\sim$100
in the HCN abundance in the inner warm envelope, consistent with the
high abundance derived from infrared absorption lines
(from: Boonman et al.\ 2001).}
\label{hcnobs}
\end{figure}

\begin{table*}[t]
\caption{Chemical characteristics of young stellar objects}
\begin{center}\scriptsize
\begin{tabular}{llllll}
\hline
\noalign{\smallskip}
Component & Chemical & Submillimeter & Infrared & Examples & Examples \\
          & characteristics & diagnostics & diagnostics & High-mass& Low-mass\\
\hline
\noalign{\smallskip}
Surrounding cloud & Low-T chemistry &Ions, long-chains &Simple ices & SgrB2 (NW) 
    & TMC-1 \\
            &                 & (HC$_5$N, ...)   & (H$_2$O, CO)&  \\
\noalign{\smallskip}
Cold envelope & Low-T chemistry, & Simple species & Ices & N7538 IRS9, 
              & N1333 IRAS4B,\\
              & Heavy depletions  & (CS, N$_2$H$^+$) 
              &(H$_2$O, CO$_2$, CH$_3$OH) & W~33A & IRAS 16293 (out)\\
\noalign{\smallskip}
Inner warm  & Evaporation  & High T$_{\rm ex}$ & High gas/solid, High
      & GL 2591, & N1333 IRAS4A (in),\\
envelope    &      & (CH$_3$OH, H$_2$CO) & T$_{\rm ex}$, Heated ices  
                  & GL 2136 & IRAS 16293 (in)\\
       & & & (C$_2$H$_2$, H$_2$O, CO$_2$) \\
\noalign{\smallskip}
Hot core & High-T chemistry  & Complex organics & Hydrides & Orion hot core, 
   & \ \ \ ? \\
      &         & (CH$_3$OCH$_3$, CH$_3$CN, & (OH, H$_2$O) 
& SgrB2(N),G34.3 \\
&&vib.\ exc.\ mol.), high HCN && W~3(H$_2$O) \\
\noalign{\smallskip}
Outflow: & Shock chemistry, & Si- and S-species & 
   Atomic lines, Hydrides & W~3 IRS5,  & L1448-mm, \\
Direct impact      &Sputtering & (SiO, SO$_2$) &([S I], H$_2$O, H$_2$ high-J) 
  & SgrB2(M) 
   & L1157 \\
\noalign{\smallskip}
Outflow: & Evaporation & Morphology 
& Hydrides & \ \ \ ? & L1527, B5 IRS1 \\ 
Turbulent entrainment  
&& (CH$_3$OH, HCO$^+$, HCN) & (H$_2$ low-J)&& Serpens SMM1 \\ 
\noalign{\smallskip}
PDR, Compact & Photodissociation,& Ions, radicals & Ionic lines, PAHs, H$_2$
& S~140, & \ \ \ ? \\
H~II regions & Photoionization & (CN/HCN, CO$^+$) &([NeII], [CII])
& W~3 IRS4 \\
 
\hline
\end{tabular}
\end{center}
\end{table*}

The chemical variations are known to occur on small scales down to
2000 AU or less. For example, Wyrowski et al. (1999) observed significant
differences between the W3(OH) and W3(H$_2$O) positions, only 7$''$
(1500 AU) apart.  Watson et al.\ (2002) find strong CH$_3$CN emission
toward only one of the two continuum sources separated by less than
4$''$ (8000 AU) in W75N, a phenomenon also observed for other sources. 
Liu \& Blake (in prep.) mapped various molecules
in a chain of 8 UC HII regions along G9.62+0.19,
with the complex and sulfur-bearing molecules appearing only in 1 or 2
`hot cores'. Thus, care has to be taken in the interpretation of
unresolved single-dish data.

The above observations have led to an operational definition of a `hot
core' as a compact ($<$0.1 pc), warm ($T>100$~K) and dense ($n> 10^6$
cm$^{-3}$) parcel of gas characterized by rich molecular line emission
(e.g, Walmsley 1992).  It is currently not clear whether the `hot
core' is a physically distinct component or whether it is simply the
warm part of the inner envelope at a certain stage of chemical
evolution.  For massive YSOs, a compressed, dense shell may be formed
at the edge of the expanding ultra- or hyper-compact H~II region due
to the pressure from the ionized gas. Chemically, there is also a
difference between simple evaporation of ices (as observed for W3
IRS5) and the subsequent `hot core chemistry' between evaporated
species leading to more complex organic molecules and highly crowded
submillimeter line spectra (as observed for W3(H$_2$O)). In this
paper, only the latter case will be termed a `hot core'.

Comprehensive surveys of a larger sample of objects in selected
molecules are only just appearing. At submillimeter wavelengths,
Hatchell et al.\ (1998a,b) observed a set of 14 UC HII regions in
selected settings with the JCMT, with about half the sources showing
the characteristic crowded hot core spectra. Ikeda et al.\ (2001)
surveyed a few characteristic complex organic molecules toward a
number of massive YSOs and found trends with increasing dust
temperature for some, but not all, species, allowing molecules which
originate in the inner envelope from high temperature chemistry or
evaporation of grain mantles to be separated from those which occur
mostly in the colder outer envelope.

Van der Tak et al.\ (2000a) selected a sample of 14 massive young
stars for JCMT observations, most of which are bright at mid-infrared
wavelengths. The infrared bright sources do not show strong emission
from complex organic species, suggesting that they have not yet
reached the hot core stage. The data have been analyzed systematically
along the lines illustrated in Figure~\ref{models}, so that reliable
empirical abundances are available for this sample.  For some sources,
jumps in the abundances of two orders of magnitude of selected
molecules (CH$_3$OH, SO$_2$) have been found in the inner envelope
(van der Tak et al.\ 2000b, 2003).


The submillimeter data of the infrared-bright sample have been
complemented by mid-infrared surveys with the ISO-SWS.  The abundances
of species like HCN and C$_2$H$_2$ (Lahuis \& van Dishoeck 2000),
CO$_2$ and H$_2$O (Boonman et al.\ 2003b,c) show systematic increases,
which correlate well with other temperature tracers such as the
far-infrared color, the gas/solid ratios and the heating signatures in
the ice profiles (e.g., Gerakines et al.\ 1999, Boogert et al.\ 2000).
Sulfur-bearing species present a more confusing picture, both at
infrared (Keane et al.\ 2001) and submillimeter wavelengths (van der
Tak et al.\ 2003), with both hot core chemistry and shocks likely
playing a role.

If higher temperatures are indeed due to increased
dispersion of the envelope (i.e., decreasing $M_{\rm env}$/$L_*$),
this `global warming' may be interpreted as an evolutionary effect and
the above features can then be used as powerful tracers of the
earliest stages of massive star formation, even before the hot core
and UC HII phases. In this respect, the analysis of massive YSOs
appears ahead of that of low-mass YSOs, where systematic trends {\it
within} the equivalent Class 0 or I stages are still difficult to
discern, partly due to a lack of infrared spectroscopy (see \S 4).


\subsection{AFGL 2591 as an example}

The case of AFGL 2591 presents a good example of the power of the
analysis techniques and the information that can be derived on the
chemical structure, both from the `empirical' and `full chemical
modeling' approaches. This source has a luminosity of $2\times 10^4$
L$_{\odot}$, and is relatively nearby ($d\approx 1$~kpc) and isolated.
Figure~\ref{gl2591} shows the best-fitting temperature and density
structure, derived from the CS line ratios (van der Tak et al.\
1999). The different chemical regimes are indicated. In the cold outer
envelope ($<$90~K), significant freeze-out occurs and the abundances
are expected to be low. In the inner envelope, ice evaporation
dominates the chemistry.  The temperatures at which ices evaporate
differ from species to species and range from as low as 20~K (pure CO
ice) to 80--100~K (CH$_3$OH and H$_2$O) (e.g., Sandford \& Allamandola
1993, Fraser et al.\ 2001).  At even higher temperatures, most of the
oxygen is driven into H$_2$O and high abundances of species like HCN
occur.

Figure~\ref{hcnobs} shows the HCN $J$=9--8 line in the $\nu_2$=0 and 1
states detected with the MPIfR/SRON 800 GHz receiver on the JCMT
(Boonman et al.\ 2001). The low-$J$ HCN $J$=1--0 through 4--3 lines
give an abundance with respect to H$_2$ of only $10^{-8}$ in the outer
envelope. The HCN $\nu_2$=1--0 infrared absorption lines indicate a
value of $\sim 10^{-6}$ (Lahuis \& van Dishoeck 2000). A similarly
high abundance is needed to fit the $\nu_2=1$ $J$=9--8 data.
Together, the data give irrefutable evidence for an abundance jump of
two orders of magnitude between the outer and inner regions.

A second example of the `empirical' modeling is provided by the analysis
of the H$_2$O data. Strong infrared absorption within the $\nu_2$=1--0
band at 6~$\mu$m has been detected with the ISO-SWS by Helmich et al.\
(1996). H$_2$O emission in the $1_{10}-1_{01}$ line at 557 GHz with
SWAS is, however, weak (Snell et al.\ 1999).  Boonman et al.\ (2003d)
analyzed the combined ISO-SWS, ISO-LWS and SWAS data and tested
different H$_2$O abundance profiles. The data are only consistent with a low
abundance of $\sim 10^{-8}$ in the outer region caused by freeze out, and
a high abundance up to $10^{-4}$ in the inner region due to evaporation.


For AFGL~2591, the `full chemical model' analysis has also been
performed. Doty et al.\ (2002) have taken the density structure of van
der Tak et al.\ (1999) and solved self-consistently for the gas
temperature. They then ran a time-dependent chemistry as a
function of position $r$ in the envelope, keeping the physical
structure fixed. The initial abundances are assumed to be low in the
outer envelope ($<$90~K), but high for selected species in the inner
envelope due to evaporation of ices. The resulting abundances $x(r,t)$
are compared with those derived from observations in the inner
and outer regions, respectively, and a best-fitting cosmic-ray
ionization rate and time is determined.   For AFGL~2591, the 
inferred young
chemical `age' of $3\times 10^4$ yr is consistent with the fact that
this source does not yet show strong radio continuum emission.

\begin{figure}[t]
\vspace{-1.2cm}
\includegraphics[width=9.0cm]{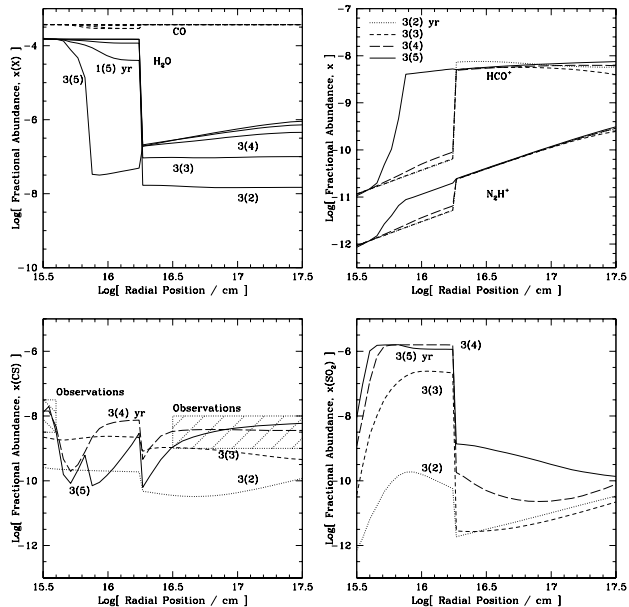}
\vspace{-3.4cm}
\caption{Examples of abundance profiles as functions of time for the
AFGL~2591 model. Top left: CO and H$_2$O; top right: HCO$^+$ and
N$_2$H$^+$; bottom left: CS; bottom right: SO$_2$.  The physical
structure is taken to be static.  For some species, a constant (e.g.,
CS), jump (e.g., H$_2$O) or anti-jump (e.g., N$_2$H$^+$) abundance
model is seen to be a good representation, whereas other molecules
vary more strongly with radius and time (from: Doty et al.\ 2002).}
\label{gl2591chem}
\end{figure}

Examples of abundance profiles as functions of time are shown in
Figure~\ref{gl2591chem}. For some species, the `jump' or `anti-jump'
model used in the empirical model is seen to be a good approximation,
but in other cases a more complex chemical profile is more
appropriate. Such detailed chemical models can be taken as guidance
for the trial abundance profiles to be used in the empirical modeling.

\section{Low-mass YSOs}

\subsection{Recent observations}

The observed low-mass YSOs are closer than their high-mass
counterparts, so that higher spatial resolution can be obtained
allowing different chemical components to be resolved. However, their
envelope masses are 1--2 orders of magnitude lower and their lines
generally weaker, especially at higher frequencies.  Many observations
have therefore been performed at lower frequencies in large beams, but
care has to be taken that such data do not apply mostly to extended,
lower-density cloud material rather than the YSO envelope. Examples of
recent chemical surveys of selected molecules toward low-mass YSOs
include Hogerheijde et al.\ (1999), Shah \& Wootten (2001), Buckle \&
Fuller (2002), J{\o}rgensen et al.\ (2002, see also this volume), and
Maret et al.\ (in preparation). While simple molecules such as HCN,
HCO$^+$, H$_2$CO and even CH$_3$OH are readily detected, none of these
sources have the crowded line spectra characteristic of high-mass hot
cores. They also do not show strong lines due to unsaturated long
carbon chains, such as seen toward some pre-stellar cores like
TMC-1S. Extremely high deuterium fractionation is observed for most
sources (e.g., van Dishoeck et al.\ 1995), including the detection of
doubly-deuterated (Loinard et al.\ 2001) and even triply-deuterated
species (van der Tak et al.\ 2002).

\begin{figure}[ht]
\includegraphics[width=5.5cm,angle=90]{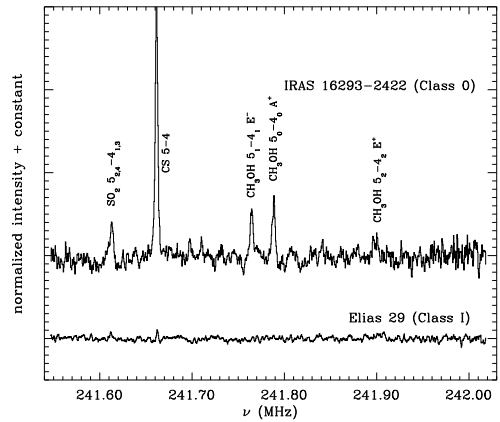}
\caption{Comparison of the emission around 241.8 GHz from the class 0
low-mass YSO IRAS 16293--2422 with that from the class I object Elias
29. Strong lines from CH$_3$OH, CS and SO$_2$ are seen in the class 0
object (from: Boogert et al.\ 2002).}
\label{i16293_e29}
\end{figure}

The sample of 16 YSOs by J{\o}rgensen et al.\ (2002) has been analyzed
systematically using the techniques outlined in Figure~\ref{models}
and covers both deeply embedded class 0 objects and more evolved class
I sources.  The figures by J{\o}rgensen et al.\ (this volume) show
some of the trends in the derived abundances. For the class 0 objects,
significant freeze-out is inferred, even for species like CO which
start to evaporate at relatively low temperatures. Class I objects
generally show very weak lines (see Figure~\ref{i16293_e29}), but the
abundances are closer to general molecular cloud values. Surprisingly
large variations in the HCN, HNC and CN abundances and abundance
ratios are found between sources (see also \S 6).  Johnstone et al.\
(2003, this volume) have observed a sample of 7 intermediate-mass YSOs
and find clear trends in the abundances with temperature of the
source.

For the class 0 objects, sufficient lines of H$_2$CO and CH$_3$OH can
be observed to test the presence of abundance jumps. Indeed, jumps of
factors of 100 or more have been inferred for several species and
sources (Ceccarelli et al.\ 2000, Sch\"oier et al.\ 2002, Maret et
al., in prep.). Such large enhancements are not only found 
in the innermost envelope, but can also occur at positions in outflow
lobes, which can be spatially separated for some low-mass YSOs (e.g.,
Bachiller \& P\'erez-Guti\'errez 1997, Garay 2000).

\begin{figure}[ht]
\includegraphics[width=8.5cm]{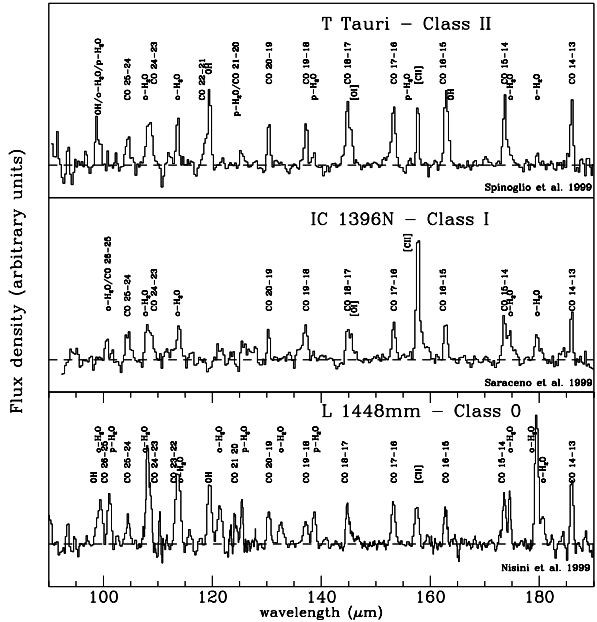}
\caption{ISO-LWS spectra of low-mass YSOs at different stages of
evolution. The continuum has been subtracted (based on Saraceno et
al.\ 1999).}
\label{isolws}
\end{figure}

At far-infrared wavelengths, 45--197 $\mu$m spectra of 28
low-luminosity YSOs have been measured with the ISO-LWS (see Nisini et
al.\ 2002 for summary). In spite of the low spectral resolution
($R\approx 200$), lines of H$_2$O, OH and CO are clearly detected in
several sources with H$_2$O strong toward class 0 objects and OH
becoming relatively more prominent toward class I objects, probably
due to enhanced UV radiation (Spinoglio et al.\ 2000,
Fig.~\ref{isolws}).  Because these data refer to large beams
($\sim$90$''$), both the envelope and large-scale outflows can
contribute to the emission, but the overall H$_2$O abundance in class
0 YSOs is high, $\sim 10^{-5}$.  For intermediate and high-mass YSOs,
the ISO-LWS spectra show much fewer H$_2$O lines (e.g., Lorenzetti et
al.\ 2002). This is partly due to photodissociation by enhanced UV
radiation, but also because these objects are more distant so that the
data suffer more from beam dilution.

Bergin et al.\ (2002) have mapped the low-mass NGC 1333 star-forming
region with SWAS in the ground-state o-H$_2$O line at 557 GHz and find
H$_2$O abundances of $> 10^{-6}$ in outflows and up to $10^{-7}$ in
quiescent gas close to luminous external heating sources.  A detailed
analysis of the ISO-LWS H$_2$O data for one object, NGC 1333 IRAS 4A,
shows low H$_2$O abundances in the outer envelope, increasing to high
values of at least $5\times 10^{-6}$ in the inner envelope (Maret et
al.\ 2002). Thus, similar H$_2$O abundance changes are seen in the
envelopes of low- and high-mass YSOs (compare AFGL~2591 case).

\subsection{IRAS 16293--2422 as an example}

The best-studied low-mass YSO in terms of chemistry is IRAS
16293--2422, a 27 L$_{\odot}$ proto-binary object located in the
Ophiuchus molecular cloud at a distance of $\sim$160 pc. It
has been coined `the low-mass equivalent of Orion-KL', although full
spectral surveys are still lacking and much fewer molecules have been
detected with the sensitivities of the observations published to date.
Blake et al.\ (1994) and van Dishoeck et al.\ (1995) performed partial
line surveys in the 230 and 345 GHz windows using matched beams with
the JCMT and CSO, respectively. Deeper integrations on selected
settings have been carried out e.g., by Ceccarelli et al.\ (2000).

\begin{figure}[ht]
\vspace{-2cm}
\includegraphics[width=11cm,angle=0]{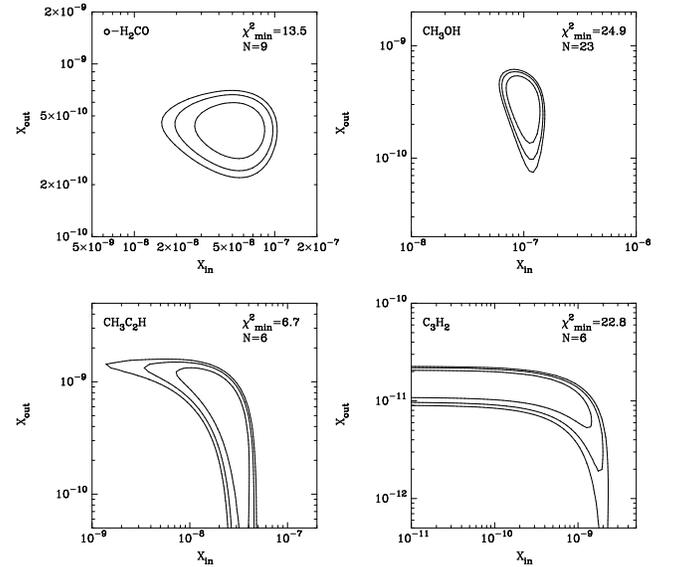}
\vspace{-6.5cm}
\caption{Examples of $\chi^2$ plots for determining the abundances of
species in the envelope of the low-mass source IRAS 16293--2422
according to the empirical model. For H$_2$CO and CH$_3$OH, many lines
covering a wide range of excitation conditions have been observed, so
that both the inner ($X_{\rm in}$)
and outer ($X_{\rm out}$) abundances are well constrained, with a
jump around 90~K. For other species, only the inner or outer abundance
is well determined
(from: Sch\"oier et al.\ 2002).}
\label{i16293_abun}
\end{figure}

Sch\"oier et al.\ (2002) have recently re-analyzed the available JCMT
and CSO data using the methods outlined in Figure~\ref{models}.  The
best-fitting temperature and power-law density profiles are very
similar to those of high-mass YSOs as shown in Figure~\ref{gl2591},
except for a scale factor. For IRAS 16293--2422, the radius at which
the dust temperature reaches 90 K is 150 AU, compared with 2500
AU for AFGL~2591.  For many species, sufficient lines covering a large
range in excitation conditions are available to fit jump models, and
$\chi^2$-fits have been used to constrain both the inner and the outer
envelope abundances. Large enhancements of factors of $\sim$100--1000
are found in the inner region above $\sim$90~K for species like
H$_2$CO, CH$_3$OH, CH$_3$CN, SO, SO$_2$ and OCS (see
Figure~\ref{i16293_abun}).  The temperature at which the jump occurs
is not well constrained but is at least 50 K.  For other molecules,
only the inner (e.g., CH$_3$C$_2$H) or outer (e.g., C$_3$H$_2$)
abundance is well determined. No complex organic molecules like
CH$_3$OCH$_3$ and HCOOCH$_3$ have been reported but the limits are not
yet very stringent compared with high-mass YSOs.

Interferometer observations of IRAS 16293-2422 reveal different
distributions for various molecules.  Molecules like CS and HCO$^+$
are good tracers of the envelope, whereas N$_2$H$^+$ and HNC clearly
avoid the inner warm region. H$_2$CO, on the other hand, traces the
inner envelope, especially in the higher excitation lines (see
Sch\"oier et al., this volume).  Such spatially-resolved data are
important to test whether the enhanced abundances in low-mass YSOs are
primarily due to thermal evaporation in the hot inner envelope or
whether grain-grain collisions in turbulent regions caused by the
interaction of the outflow with the inner envelope can also release
grain mantles.


Doty et al.\ (this conference and in prep.) have performed full
chemical modeling for this source. Starting from the (static)
temperature and density profile derived by Sch\"oier et al.\ (2002)
and using the same chemical model as applied to AFGL 2591, the
abundances of many species have been computed as functions of position
and time in the envelope.  Subsequently, the emerging line fluxes have
been obtained by calculating the excitation and radiative transfer
through the model envelope and comparing them directly with the
observational data after convolution with the appropriate beams. 
For a best-fitting time of a few $\times 10^4$ yr, the model
fluxes generally agree within 50\% with the data, much better than the
factor of 3--10 agreement generally cited for dark clouds (e.g.,
Terzieva \& Herbst 1998).

\section{Models of hot core and envelope chemistry}

\subsection{Static models}

In the static envelope or `cocoon' models described in \S 3.2 and 4.2,
the temperature and density structures are taken to be constant in
time and the molecules do not move from one position to another within
the envelope, e.g., due to infall. These 1D models can contain a `hot
core' in the center, where evaporation of ices followed by
high-temperature chemistry occurs. Static hot core models have
been extensively described by Charnley et al.\ (1992, 1995), Charnley
(1997), Caselli et al.\ (1993) and Rodgers \& Charnley (2001) (see
Millar 1997 for review). The choice of molecules to be evaporated and
their abundances are inspired by the observed chemical composition of
ices toward high-mass YSOs, which consists largely of H$_2$O, CO$_2$,
CH$_3$OH and H$_2$CO (e.g., Gibb et al.\ 2000). NH$_3$ and H$_2$S are
often assumed to represent the major nitrogen and sulfur reservoirs
in the ices, although this has not yet been confirmed by
observations (see Taban et al.\ 2003 for the case of NH$_3$). The hot
core models are particularly sensitive to the initial CH$_3$OH/NH$_3$
ratio.

In most hot core models, all ices are assumed to evaporate
instantaneously at $t$=0, but in some cases a more gradual evaporation
is used (e.g., Viti \& Williams 1999). The evaporated ices
subsequently drive a rapid high temperature chemistry in the warm gas,
leading to more complex organic molecules such as CH$_3$OCH$_3$ and
HCOOCH$_3$ on a time scale of $10^4$ yr. Nitrogen-bearing molecules
require longer timescales up to $10^5$ yr and favor higher
temperatures.  Similarly, the H$_2$S/SO$_2$ ratio can serve as a
chemical clock, if it is assumed that most of the gas-phase sulfur is
initially in H$_2$S.  Indeed, one of the main limitations in the use
of these molecules to constrain the source age is that it is not known
whether they are `first generation' molecules present in the ices or
whether they represent `second generation' species produced in the hot
gas.

The combination of a hot core with a more extended envelope was first
performed in a 3-layer model for G34.3 by Millar et al.\ (1997) and
subsequently within a single power-law model for AFGL 2591 by Doty et
al.\ (2002). In the latter case, the density and dust temperature
profiles are taken directly from observations. The gas temperature is
calculated explicitly taking all heating and cooling mechanisms of
the gas into account.  The chemistry is then solved as a function of
time at each position using an extensive chemical network containing
both low- and high-temperature reactions and separating the envelope
into a warm inner part where the ices are evaporated and a cold outer
part where molecules are still heavily depleted.

\subsection{Dynamical models}

In the dynamical models, the temperature $T$($r$,$t$) and density
$n$($r$,$t$) of the source change according to some physical
prescription and the gas moves inward or outward according to a
velocity profile $v$($r$,$t$). The best studied case is that of the
popular Shu (1977) collapse model appropriate for low-mass YSOs.
Rawlings et al.\ (1992) were the first to follow the chemistry
$x$($r$,$t$) in a parcel of gas as it is falling into the center in an
effort to explain the difference in chemistry and line profiles for
species such as NH$_3$ and CS. Ceccarelli et al.\ (1996) subsequently
computed the full thermal balance of the dust and gas as a function of
time, coupled with a simple chemistry aimed at predicting the [O I]
and H$_2$O emission from low-mass YSOs. More recently, Rodgers \&
Charnley (in prep.) coupled a full chemical network with a Shu-type
dynamical model and compared the results with those of a static
model. Interestingly, they concluded that small-scale differentiation
and high abundances of complex species can only occur if the gas is
not falling rapidly into the protostar, i.e., observations of such
species require a more or less static model. In a dynamical model, the
chemical timescales become longer than the dynamical timescales so
that the abundances are frozen at their initial values.

\begin{figure}[ht]
\includegraphics[width=7cm,angle=0]{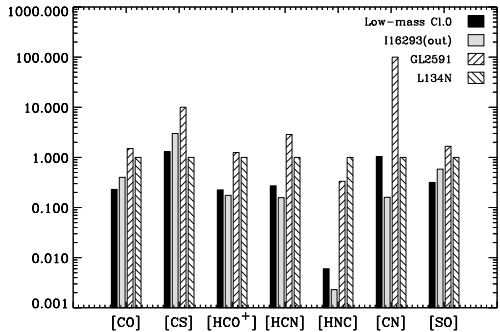}
\caption{Comparison of the abundances found in the cold outer envelope
of the low-mass YSO IRAS 16293-2422 (Sch\"oier et al.\ 2002) with
those derived for a larger set of class 0 low-mass YSOs (J{\o}rgensen
et al., in prep) and the high-mass AFGL~2591 envelope (van der Tak et
al.\ 2000a, 2003).  Note the strong variations in the HCN/HNC/CN
abundances and ratios (from: J{\o}rgensen et al., in
prep.).}
\label{abun_cold}
\end{figure}

\section{High versus low-mass YSOs}

It is clear from the preceding sections that there are several
similarities but also some potential differences between the chemistry
in low- and high-mass YSOs. Since statistically significant data sets
analyzed in a homogeneous way are still lacking, some of the apparent
differences may well become similarities and vice versa in future
studies.

\subsection{Similarities?}

{\it Physical structure:} The density structure of the envelopes of
both low- and high-mass YSO's can be well described by a smooth power
law whose index can be constrained observationally from dust continuum
and/or line observations. Some analyses suggest a slightly shallower
index for high-mass YSOs compared with low-mass YSOs (e.g., van der
Tak et al.\ 2000a), whereas other studies find similar indices within
the observational errors (e.g., Mueller et al.\ 2002, Beuther et al.\
2002), but this potential difference does not have significant
chemical consequences to first order. In both sets of sources, the
dust temperature follows an optically thin $\propto r^{-0.4}$ law in
the outer envelope, increasing more steeply in the inner part to
temperatures above 100~K. The only difference is a scale factor,
resulting in a larger physical size of the warm region for high-mass
YSOs. The good fits to the data imply that outflows and shocks give
only a limited contribution to the heating in both types of objects.
Because of the higher temperatures throughout, the abundances in the
outer envelopes of high-mass objects may be somewhat higher than those
in low-mass class 0 YSOs (see Fig.~\ref{abun_cold}).

\begin{figure}[ht]
\includegraphics[width=7cm,angle=0]{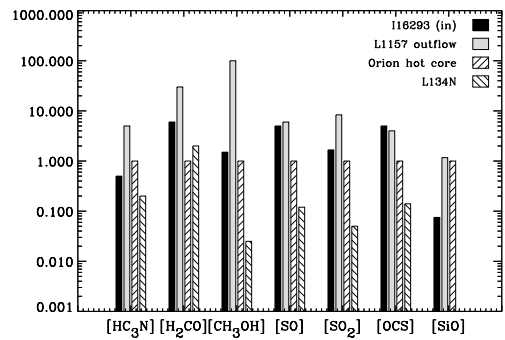}
\caption{Comparison of the abundances found in the inner warm region
of the low-mass YSO IRAS 16293-2422 with those derived for the
high-mass Orion hot core and for the outflow position offset from the
low-mass YSO L1157 The abundances are normalized to the Orion hot
core (based on:
Sch\"oier et al.\ 2002).  }
\label{abun_warm}
\end{figure}

{\it Abundance jumps:} For both low- and high-mass YSOs, jumps in the
abundances of selected molecules are found (e.g., CH$_3$OH,
H$_2$O). The jumps are of comparable magnitude (about a factor of 100)
and their location is consistent with the temperature of $\sim$90~K at
which most ices are thought to evaporate.  Figure~\ref{abun_warm}
compares the abundances of a few characteristic species in the Orion-KL
hot core with those found in the inner warm region of IRAS 16293--2422
(see Table 7 of Sch\"oier et al.\ 2002 for details). The number of
objects for which such detailed abundance analyses have been done is
still small however, and statistics on larger samples are needed.

\subsection{Differences}

{\it Hot core chemistry:} Both low- and high-mass YSOs are known to
have an inner warm and dense region where ices have evaporated. But do
they have a similar hot core chemistry? Complex organic molecules such
as CH$_3$OCH$_3$ and HCOOCH$_3$, which are thought to be
secondary species produced by high-temperature reactions, are 
found toward high-mass YSOs only. These species have not yet been
detected toward low-mass YSOs in published spectra, although the
abundance limits are comparable to those in high-mass YSOs if the
complex molecules are located only in the inner  warm
region. Deeper searches are needed for low-mass YSOs.  Potential
differences in hot-core chemistry could be due to several
factors. First, the dynamical infall timescales in the warm gas in
low-mass YSOs are much shorter ($<10^3$ yr)
compared with high-mass YSOs, so that the complex chemistry may
not have enough time to develop unless the collapse is slowed
down. Second, the chemical composition of the ices may be different
due to different H/H$_2$ ratios in the collapsing envelope and/or
different grain temperatures in high- versus low-mass star-forming
regions. Third, the size of the hot core and the temperature region
above 230~K may be so small in low-mass YSOs ($<$100 AU, i.e.,
comparable to the size of the disk) that it is negligible in
practice. On the other hand, expansion of the ultracompact H~II region
may create a large compressed shell of warm molecular gas for
high-mass YSOs.

{\it Deuterium fractionation:} The observed ratios of deuterated
molecules with respect to their normal counterparts are at least an
order of magnitude higher on average for low-mass YSOs (e.g., Roberts
et al.\ 2002). This can be due to two factors. First, the temperature
of the pre-stellar cores may be lower for low-mass YSOs, enhancing the
deuterium fractionation especially if CO is frozen out. In some
high-mass star-forming regions, the temperatures may never have been
cold enough for significant CO depletion. Second, the time-scales of
the pre-stellar phase may have been longer for low-mass YSOs, leading
to more complete CO freeze-out across the core and further favoring
the production of H$_2$D$^+$.

{\it Effects of UV or X-rays:} For both low- and high-mass YSOs no
clear diagnostics of UV radiation or X-rays have yet been found in the
deeply embedded phase. The UV radiation can be either externally or
internally produced. In the former case, the dust and gas temperatures
in the outer envelope are raised and molecules are photodissociated.
For example, HCN is photodissociated to CN, a well-known
photodissociation product in PDRs (e.g., Fuente et al.\ 1998). The
internal UV radiation or X-rays can originate either from the young
star itself or from a hot accretion boundary layer between the inner
disk and the star. Characteristic products are CO$^+$, CN and OH.
Moreover, the radiation can affect the composition of the ices, e.g.,
by producing a feature of OCN$^-$ at 4.62 $\mu$m (but see van
Broekhuijzen et al., this volume).  UV radiation also excites PAH
features. Deep searches for photodissociation and X-ray products and
PAH features in both low- and high-mass YSOs are needed. It is clear
that UV radiation plays an important role in the chemistry of
high-mass YSOs in later stages during the evolution from the
ultracompact to the classical H~II region stage, where PAHs and
photodissociation products are widely observed.

{\it Effect of shocks:} Outflows are known to be associated with both
low- and high-mass YSOs and their effects on the chemistry are well
documented at {\it off-source} positions in the outflow lobes, where
enhancements of species like SiO, CH$_3$OH and H$_2$O have been seen
(see also Figure~\ref{abun_warm}). However, their importance in the
inner dense envelope relative to thermal evaporation remains to be
determined. The observed line widths toward both low- and high-mass
YSOs are typically a few km s$^{-1}$ and only the strongest lines of a
few species show clear evidence for line wings (see also Buckle \&
Fuller 2002).  It has been argued that shocks may be relatively more
important in low-mass YSOs, not only for the dispersion of the
envelope but also for releasing grain mantles in the turbulent shear
zones between the outflow and the quiescent envelope. This can be
tested by mapping the distribution of grain-surface products such as
H$_2$CO. High-angular resolution observations of a few low-mass
sources provide evidence for both scenarios (e.g., Sch\"oier et al.,
in prep.).  Schreyer et al.\ (2002) show that the presence of low-mass
outflows within a high-mass envelope does not have noticeable effects
on the chemistry, indicating that only a small fraction of the gas is
affected. On the other hand, the absence of abundant gas-phase CO$_2$
in the inner envelope has been cited as evidence for shock chemistry
throughout high-mass YSO envelopes (Charnley \& Kaufman
2000). Hatchell et al.\ (2001) argued based on SiO data that shocks do
not affect the hot core of the massive YSO G34.3, although they are
present in the more extended envelope. Further observations of shock
diagnostics, such as the NS/CS ratio proposed by Viti et al.\ (2001),
are needed (see also Hatchell \& Viti 2002).

{\it Geometry:} The effects of geometry and foreground clouds are
thought to be more important for the analysis of low-mass YSO data
than for high-mass sources, where the envelope often overwhelms all
other emission or absorption. For example, a significant fraction of
the ices along the line of sight toward low-mass YSOs in Ophiuchus may
originate in unrelated foreground material (e.g., Boogert et al.\
2002).  Low-excitation millimeter lines often have their origin
primarily in the surrounding clouds, where they may or may not be
distinguished by their velocity structure.  Finally, disks are known
to exist in the embedded class I phase for low-mass YSOs, whereas
their presence for high-mass YSOs remains to be firmly
established. Depending on the viewing angle, disks can change the
observed mid-infrared characteristics of the source. They also provide
a natural cut-off for (part of) the inner warm envelope and may shield
some of the envelope from heating. Their chemical structure is
expected to be significantly different from that of the inner envelope
due to the high densities and low mid-plane temperatures.


\section{Conclusions}

The combination of submillimeter and infrared observations allows a
nearly complete inventory of gas- and solid-state species of high- and
low-mass YSOs. Combined with detailed modeling using a new generation
of flexible radiative transfer codes, quantitative analyses of both
their physical and chemical structure are now possible. Clear
variations in the gas-phase abundances and the gas/solid ratios are
found with increasing temperature in the envelope, indicating that the
chemistry is controlled by freeze-out and ice evaporation. Systematic
studies of statistically significant samples of low- and
high-mass YSOs are only just appearing, and some similarities and
differences are emerging. Due to the limited samples and poor
sensitivity and angular resolution of the existing data, much more
work is needed to draw firm conclusions.

Progress in this field will be greatly aided by new observational
facilities. SIRTF and eventually JWST will allow mid-infrared imaging
and spectroscopy of a much larger sample of embedded low-mass YSOs.
Herschel will provide unique data on H$_2$O, OH and other hydrides in
both low- and high-mass YSOs, which, among many other applications,
will help to disentangle the importance of shocks and thermal
evaporation.  Finally, ALMA will give maps of many species at
unprecented angular resolution and sensitivity, essential to constrain
the small-scale structure of the envelope and outflows and probe the
circumstellar disks. Together, they will allow the full chemical
history of material from molecular clouds to planetary systems to be
traced.

\begin{acknowledgements}

The author is grateful to Geoff Blake, Adwin Boogert, Annemieke
Boonman, Rogier Braakman, Steven Doty, Helen Fraser, Michiel
Hogerheijde, Doug Johnstone, Jes J{\o}rgensen, Fred Lahuis, Klaus
Pontoppidan, Fredrik Sch\"oier, Willem Schutte, Fleur van Broekhuizen
and Floris van der Tak for many enjoyable discussions and for
providing input and figures for this review. The reader is strongly
encouraged to read their original papers on these topics, some of
which are summarized in this volume.  Astrochemistry in Leiden is
supported by grants no.\ 614.41.003 and 614.041.004 as well as a
Spinoza grant from NWO and by the Netherlands Research School of
Astronomy (NOVA).

\end{acknowledgements}

\end{document}